\documentclass[a4paper,twoside]{article}

\usepackage{arxiv}

\usepackage[utf8]{inputenc}
\usepackage[T1]{fontenc}
\usepackage[hidelinks]{hyperref}
\usepackage{url}
\usepackage{booktabs}
\usepackage{amsfonts}
\usepackage{nicefrac}
\usepackage{microtype}

\usepackage{graphicx}
\usepackage[numbers]{natbib}
\usepackage{doi}

\usepackage{subcaption}
\usepackage{calc}
\usepackage{amssymb}
\usepackage{amstext}
\usepackage{amsmath}
\usepackage{amsthm}
\usepackage{multicol}
\usepackage{pslatex}
\usepackage{algorithm2e}
\usepackage[style=english]{csquotes}
\usepackage{enumerate}
\usepackage[bottom]{footmisc}
\usepackage{cleveref}

\title{Enhancing Trust in Inter-Organisational Data Sharing: Levels of Assurance for Data Trustworthiness}

\date{}

\usepackage{authblk}

\newbox{\orcid}\sbox{\orcid}{\includegraphics[scale=0.06]{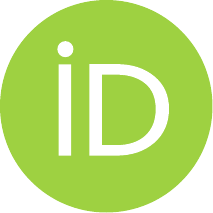}} 
\author[1]{%
	\href{https://orcid.org/0009-0002-8060-7162}{\usebox{\orcid}\hspace{1mm}Florian~Zimmer}%
}
\author[2]{%
	\href{https://orcid.org/0000-0001-5494-9770}{\usebox{\orcid}\hspace{1mm}Janosch~Haber}%
}
\author[3]{%
	\href{https://orcid.org/0000-0001-9873-2557}{\usebox{\orcid}\hspace{1mm}Mayuko~Kaneko}%
}
\affil[1]{Fraunhofer Institute for Software and Systems Engineering ISST, Speicherstraße 6, Dortmund, 44147, Germany}
\affil[2]{Fujitsu Research of Europe, 9 Albert St, Slough SL1 2BE, United Kingdom}
\affil[3]{Fujitsu Limited, 4-1-1 Kamikodanaka, Nakahara-ku, Kawasaki, Kanagawa 211-8588, Japan}

\renewcommand{\shorttitle}{Levels of Assurance for Data Trustworthiness}

\hypersetup{
	pdftitle={Enhancing Trust in Inter-Organisational Data Sharing: Levels of Assurance for Data Trustworthiness},
	pdfauthor={Florian~Zimmer, Janosch~Haber, Mayuko~Kaneko},
	pdfkeywords={Data Trustworthiness, Levels of Assurance, Inter-Organisational Data Sharing, Trust, Data Spaces, Design Science Research},
}

\begin{document}

\maketitle

\begin{abstract}
    As data is increasingly acknowledged as a highly valuable asset, much effort has been put into investigating inter-organisational data sharing, aiming at utilising the value of formerly unused data. Moreover, most researchers agree, that trust between actors is key for successful data sharing activities. However, existing research oftentimes focus on trust from a data provider perspective. Therefore, our work highlights the unbalanced view of trust, addressing it from a data consumer perspective. More specifically, our aim is to investigate trust enhancing measures on a data level, that is data trustworthiness. We found, that existing data trustworthiness enhancing solutions do not meet the requirements of the domain of inter-organisational data sharing. Therefore, our study addresses this gap. Conducting a rigorous design science research approach, this work proposes a new Levels of Assurance for Data Trustworthiness artifact. Built on existing artifacts, we demonstrate, how it addresses the identified challenges within the domain appropriately. We found that our novel approach requires more work to be suitable for adoption. Still, we are confident that our solution can increase consumer trust. We conclude by contributing to the body of design knowledge and emphasise the need for more attention to be put into consumer trust.
\end{abstract}

\keywords{Data Trustworthiness, Levels of Assurance, Inter-Organisational Data Sharing, Trust, Data Spaces, Design Science Research.}

\section{Introduction}
\label{sec:introduction}

The increasing adoption of information driven technology across industries, and its integration in nearly every aspect of life, are highlighting the ever-growing importance of data. Data is considered a central driver in the acceleration of digital transformation \cite{Otto.2022b}. Consequently, the \textit{European Data Strategy}\footnote{\url{https://commission.europa.eu/strategy-and-policy/priorities-2019-2024/europe-fit-digital-age/european-data-strategy\_en} [Accessed 30.03.2025]} was proposed by the European Commission in 2020, acknowledging data as an essential asset for innovation and growth in businesses and societies as a whole. As a result, inter-organisational data sharing has gained much attention in academia and industry recently, aiming to unlock the full potential of previously unused and under-utilised data \cite{Tocco.2022}.

However, organisations are often  hesitant when engaging in data sharing activities. A number of barriers prevent a more wide-spread adoption of inter-organisational data sharing, with a lack of trust and transparency between actors mentioned as one of the most fundamental barriers \cite{Jussen.2023}. A main source for the lack of trust in inter-organisational data sharing, mentioned in research, are challenges to \textit{data sovereignty}, i.e., the concern of data providers to lose control over their data once shared with other organisations \cite{Scherenberg.2024}.

In order to address these concerns, the concept of \textit{data spaces} has emerged to provide a framework for data sharing, establishing data sovereignty, by enabling mutual trust in the interoperable data exchange between data providers and consumers \cite{Otto.2022b}. However, the issue of trust here is predominantly considered from the perspective of the data provider. We found, that approaches often aim at preventing the loss of corporate knowledge or sensitive information \cite{DHauwers.2022,Huber.2022}, or the security of the data exchange itself \cite{Huber.2022,Hompel.2022} - but rarely mention the consumer's risks and their need for trust in data providers and the data made available by them \cite{Otto.2022b,Tocco.2022}.

Contrarily, in the ISO/IEC 15408-1 standard for information security, data, or more specifically information consumers are also referred to as \textit{risk owners}, as they are confronted with the potential risks resulting from data providers taking insufficient measures to assure data integrity \cite{InternationalOrganizationforStandardizationISOInternationalElectrotechnicalCommission.August2022}. Because data is becoming so essential for decision making \cite{FaheemZafar.2017}, leveraging  (un-)intentionally modified, incomplete, or compromised data exposes data consumers to potentially severe consequences from financial losses up to human harm \cite{Gomez.2009}. Still, data consumers usually have no other option than to trust data providers, as data trustworthiness cannot be established otherwise \cite{Alhaqbani.2009,Tao.2020}. 

In this paper, we argue that trust in inter-organisational data sharing should not be limited to the organisational level but must also encompass the trustworthiness of the data itself. Following a rigorous \textit{design science research} (DSR) approach, we review the literature on data trustworthiness across various domains and identify challenges in ensuring it in complex sharing scenarios.

As a result, we adopt a multi-dimensional view of the problem and solution space and develop a novel artifact, aiming at addressing the shortcomings of existing artifacts. We introduce the groundbreaking concept of \textit{Levels of Assurance for Data Trustworthiness}, or \textit{Data LoA}, which offers new means of enhancing trust and transparency among data providers and data consumers - including additional data assurances to enhance trust for data consumers. In this first iteration, we aim to build a foundational model, outlining key actors and their interactions. Furthermore, an instantiating artifact is presented in the form of a \textit{proof of concept} (PoC) implementation, demonstrating the LoA concept from a technical data sharing perspective. Our main contributions include:

\begin{enumerate}[(i)]
    \item Compiled design knowledge, identifying existing work and mapping the problem and solution space
    \item A novel Data LoA artifact, aiming at enhancing data consumer trust
    \item An experimental proof of concept, showcasing the practical application of the Data LoA
\end{enumerate}

The remainder of this paper is structured as follows: in Section~\ref{sec:related_work}, we touch upon relevant related work. Section~\ref{sec:methodology} outlines the research methodology followed in this study. Section~\ref{sec:results} presents the design knowledge and objectives derived from the literature review, and proposes the concept of Data LoA as an artifact. In Section~\ref{sec:discussion}, we discuss implications as well as limitations and touch points for future work. Section~\ref{sec:conclusion} concludes the paper with a brief summary.

\section{Related Work}
\label{sec:related_work}

\subsection{Data Trustworthiness}
The trustworthiness of data has been extensively studied across various domains and applications such as healthcare, defence, traffic control, and manufacturing \cite{Gomez.2009}. The main concentration of work on data trustworthiness however is in the contexts of \textit{internet of things (IoT)} and \textit{mobile crowd sensing (MCS)} - both areas, that face the challenges of an unknown, potentially insecure network of known and unknown actors.

Data trustworthiness is usually described as the possibility to ascertain the correctness of data provided by a data source \cite{Haron.2017}. Yet, a high degree of context and domain dependency have prevented the formulation of a generally accepted notion of data trustworthiness so far \cite{Bertino.2015,Rahman.2019,Xu.2023}. Circumnavigating a holistic definition, literature oftentimes mentions and focuses on specific dimensions of data trustworthiness; most commonly data availability, quality, security, and compatibility - each of which concerned with a number of different aspects themselves \cite{Xu.2023}. 

With many different approaches to assure, measure, and define data trustworthiness existing, \cite{Junior.2021} proposed a taxonomy for artifacts in IoT. The taxonomy mainly groups them by conceptual trust definitions, mathematical models, or trust frameworks. For instance, \cite{Ormazabal.2024} present a trust assessment canvas to gauge the trustworthiness of publicly available medical data. \cite{Foidl.2023} on the other hand, present a trust score model capable of measuring the trustworthiness of industrial IoT data sources. Furthermore, \cite{Leteane.2024} propose a trust enhancing framework for data traceability in the context of food supply chains. 

Based on this, some researchers argue that the variety of different approaches and notions, prevents a development of a comprehensive solution to assure data trustworthiness \cite{Bertino.2015,Ebrahimi.2022}. Therefore, \cite{Haron.2017} argues that a combination of different techniques is required, in order to meet the requirements of data consumers, highlighting the need for overarching solutions.

\subsection{Levels of Assurance}
\textit{Levels of Assurance} (LoA) refer to the degree of confidence that can be assigned to some kind of entity, process, or system acting or operating as claimed \cite{InternationalOrganizationforStandardizationISOInternationalElectrotechnicalCommission.April2013}. LoAs are an assurance technique, used to evaluate and grade complex scenarios, simplifying and improving decision making and risk management \cite{Nenadic.2007}. More formally, a \textit{relying party} utilises provided LoA information to determine their level of confidence in the credibility of a \textit{claimant's} claim. Usually, there is at least one other party involved, namely the \textit{assurance provider}, which audits and assures the claimant's claim \cite{MartinezFerrero.2018,InternationalOrganizationforStandardizationISOInternationalElectrotechnicalCommission.April2013}. If there is no external assurance provider, claims are self-asserted.

LoAs are mainly used in the domain of identity validation, for example in the ISO/IEC 29115\footnote{\url{https://iso.org/standard/45138.html} [Accessed 30.03.2025]} standard, the eIDAS\footnote{\url{https://digital-strategy.ec.europa.eu/en/policies/eidas-regulation} [Accessed 30.03.2025]} regulation as proposed by the European Commission, or the NIST 800-63-A\footnote{\url{https://pages.nist.gov/800-63-3/sp800-63a.html} [Accessed 30.03.2025]} guidelines. The concrete LoA levels define the processes, management activities, and technologies needed in order to establish different levels of confidence in the claimed identity.
Nonetheless, LoAs are also adopted widely within other domains, such as the ISO 26262-3\footnote{\url{https://iso.org/standard/68385.html} [Accessed 30.03.2025]} standard for automotive safety or the CSA STAR\footnote{\url{https://cloudsecurityalliance.org/star} [Accessed 30.03.2025]} program for cloud security.

LoAs are usually risk-based, defining which dimensions of risk must be addressed and mitigated in order to assure the credibility of a claim. Hence, the higher the perceived risk for the relying party, the higher the required level of confidence in the claim's validity, and thus the LoA should be \cite{MartinezFerrero.2018,InternationalOrganizationforStandardizationISOInternationalElectrotechnicalCommission.April2013}. A comprehensive LoA concept should guide the claimant on how to mitigate risks, while providing the relying party with assurances needed for informed decision making.

Besides providing risk management capabilities, the eIDAS regulation also mentions improving trust among adopters. They achieve this by providing assurances of audited identification techniques and clearly defining liabilities to specify each party's responsibilities. Therefore, commonly used identification techniques are certified at different LoAs within eIDAS. This provides standardised assurances relevant for assessments of mutual trust, that enable interoperability in the heterogeneous identification techniques landscape of the EU \cite{EuropeanParliament.23July2014}.

\section{Methodology}
\label{sec:methodology}
This paper aims to address the lack of trust enhancing means for data consumers in inter-organisational data sharing. More specifically, we address the assurance of the trustworthiness of data by designing a new artifact for \textit{Levels of Assurance for Data Trustworthiness} (Data LoA). The ultimate goal of this artifact is twofold: on the one hand, it aims to provide data consumers with risk assessment capabilities. On the other, it equips data providers with standardised principles on how to establish different degrees of data trustworthiness assurances. Together, these mechanisms are aimed at enhancing trust in inter-organisational data sharing and enabling interoperability among existing trust assuring solutions.

To develop our artifact, we conducted a rigorous DSR approach, following \cite{Peffers.2007}. More specifically, we followed an objective-centred approach. This is based on our finding, that a range of data trustworthiness assuring artifacts already exist. Yet, they fail to provide comprehensive means of assuring trust in the broader context of inter-organisational data sharing in a standardised, interoperable manner.

\begin{figure}[!h]
  \centering
      {\includegraphics[width = 4.5cm]{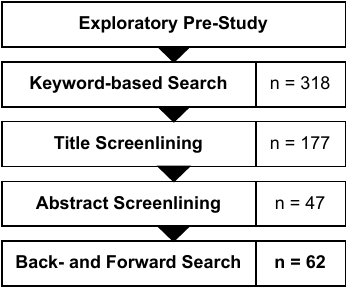}}
  \caption{Conducted literature search approach.}
  \label{fig:slr-approach}
 \end{figure}

Furthermore, we found that previous artifacts were not necessarily developed following DSR. Thus, current literature does not provide a sound foundation of design knowledge to build upon on. We therefore started by identifying the relevant problem and solution space, specifying the challenges, motivations, and goals addressed by existing artifacts. We did so by following a rigorous \textit{structured literature review (SLR)} as described by \cite{vomBrocke.2015}, deriving design knowledge by empirical means. According to the authors, literature reviews are a proven method on justifying the novelty of design ideas, as well as informing the design process. The SLR process is pictured in Figure~\ref{fig:slr-approach}.

Therefore, as suggested by \cite{vomBrocke.2015}, we first started with an unstructured exploration of relevant literature, mainly using Google Scholar, in order to increase the familiarity with the subject. Then, we continued with a structured keyword search in the IEEExplore, ACM, and ScienceDirect databases. These resources were chosen due to their importance in the first stage, as well as their relevance to related subjects, and coverage of adjacent fields. We used the search term \textit{("Data" AND ("Trustworthiness" OR "Trustworthy"))}, matching titles in research articles only to narrow down the scope and improve the relevance of matches. We executed the search in December 2024, which returned 318 matches. 148 of these came from IEEExplore, 121 from ACM, and 49 from ScienceDirect. We screenlined these articles' titles and removed unequivocal false positives, which resulted in 177 articles that were more thoroughly investigated.

We then excluded articles that were not related to measuring, evaluating, or assuring the trustworthiness of data. This resulted in a total of 47 articles. Using this collection, we conducted backward and forward searches by thoroughly reading each article, looking for key references, and using Google Scholar to execute a targeted forward search. According to \cite{vomBrocke.2015} back- and forward searches are a good way to ensure sufficient coverage of relevant literature. Doing this left us with an additional 35 articles, 20 of which were already part of our set of identified literature. Thus, we ended up with a total of 62 articles to consider.

After analysis of these articles, we empirically derived design knowledge for two topics: i) challenges and motivations for means to measure or assure data trustworthiness, and ii) common objectives of existing artifacts. We did this in order to ground the relevancy of our to be designed artifact and guide our develop and design efforts by goals often followed. Doing this provides a strong foundation to build a new artifact on top of, facilitating existing research, as also highly suggested by \cite{vomBrocke.2020}.

\begin{figure}[!h]
  \vspace{-0.2cm}
  \centering
      {\includegraphics[width = 7.5cm]{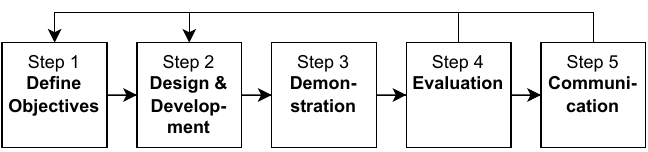}}
  \caption{The objective-centred DSR approach following \cite{Peffers.2007} applied in this study.}
  \label{fig:dsr-approach}
 \end{figure}

We continued our DSR objective-centred approach as pictured in Figure~\ref{fig:dsr-approach} by posing the leading question, as suggested by \cite{Peffers.2007}: \textit{What would a better artifact accomplish?} To address this question, we derived and selected a set of design objectives aligned with our overall goal of enhancing data trustworthiness in inter-organisational data sharing, and used these objectives to guide the development of our artifact. As we have been already familiar with the concept of LoAs for identity validation, it quickly became evident that LoAs serve a similar purpose. Thus, we used them as foundation to build upon on, adjusting the design goals accordingly.

Based on this, a cross-functional team of researchers and practitioners developed the Data LoA artifact. Doing this, we mainly focused on defining central actors as well as their relations, in order to establish a sound foundation for future iterations to build on top of. We therefore did not define specific levels yet, as a lack of a common notion of data trustworthiness made it hard to specify them at this early stage.

We then proceeded with evaluating our artifact by instantiating a PoC in the context of data spaces, which was deemed to be a relevant and informative domain for investigating trust effects in inter-organisational data sharing. By conducting an experimental simulation, the PoC allowed us to assess the technical feasibility of our concept and determine limitations and considerations for future work. The communication step, as proposed by \cite{Peffers.2007}, is performed by publishing this paper.

\section{Results}
\label{sec:results}
The aim of this paper is to create a new trust enhancing technique that addresses the lack of trust from a consumer perspective in inter-organisational data sharing. To do so, we adopted an objective-centred DSR approach to develop an artifact based on design knowledge of existing literature. Therefore, we first identify both problem and solution space by conducting a literature review in order to derive such design knowledge, following \cite{vomBrocke.2020}. Based on this, we derive design objectives stipulated by existing solutions, and describe and demonstrate our resulting artifact.

\subsection{Design Knowledge \& Grounding}
\begin{figure}[!h]
  \vspace{-0.2cm}
  \centering
      {\includegraphics[width = 7.5cm]{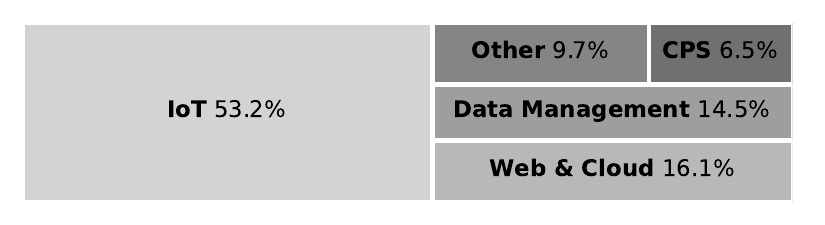}}
  \caption{ The domain distribution of the 62 investigated articles. }
  \label{fig:domain-distribution}
 \end{figure}

Adopting an objective-centred approach comes with the assumption, that the problem space is already well-defined and investigated in literature \cite{Peffers.2007}. Therefore, we conducted a SLR in order to extract and derive existing design knowledge, in order to build our artifact upon on. The domain distribution of the 62 investigated articles can be seen in Figure \ref{fig:domain-distribution}. It is apparent that most of the data trustworthiness related studies were conducted in the broader area of IoT, as well as Web \& Cloud. The commonalities in regards to the problem space, thus acknowledged problems and motivation, are described in the following. For brevity, only key literature is referenced. A full overview of considered articles, as well as a detailed overview of the derived clusters can be found in \cite{anonymous.2025}.

\textbf{Undeterminable Risk.} A major issue and motivation for enhanced trust in data are undeterminable risks. Most studies agree that consuming untrustworthy data, i.e., data for which trustworthiness is not assured, is linked to great risks. According to \cite{Haron.2017,Karthik.2016} this is because data is usually utilised for some form of decision making. The reliability and accuracy of this decision making therefore is greatly influenced by the data that it is grounded on. Consequently, using untrustworthy data can lead to severe consequences - as incidents in healthcare or power supply have previously demonstrated \cite{Jaigirdar.2019,Lim.2012}. Ensuring data trustworthiness is therefore primarily motivated by better risk management capabilities, where, e.g., low trustworthy data will only be used for low risk applications.
    
\textbf{Challenging Complexity.} Many previous papers argue that assuring and measuring data trustworthiness is a complex and challenging task. According to \cite{Bertino.2015}, e.g., addressing the different facets of data trustworthiness also requires a complex combination of different approaches and techniques. \cite{Haron.2017} further emphasise this by reflecting on the complexity added by heterogeneous IoT environments, and \cite{Prandi.2015} recognise data trustworthiness in the context of MCS as key challenge. What is more, data trustworthiness measuring techniques often incorporate an array of individual metrics to compile a final trust score. In a concrete example, \cite{Leteane.2024b} employ multiple trust metrics such as device malfunction, data tampering, and battery problems to measure and determine trustworthiness.

\textbf{Increasing Demand.} At different points in time, studies have emphasised the growing need for trustworthy data as well as assurance and quantification means. For instance, \cite{Bertino.2015} states that \textquote{[...] today’s demand for data trustworthiness is stronger than ever. As many organizations are increasing their reliance on data for daily operations and critical decision making, data trustworthiness is arguably one of the most critical issues.} More recently, \cite{Islam.2025} suggest trust estimation of data as one way to tackle the growing challenges of data related risks and vulnerabilities, and \cite{Anjomshoaa.2022} highlight the growing need for trustworthy data for use in \textit{artificial intelligence} (AI).

\textbf{Success Factor.} Another widely acknowledged fact in literature is, that establishing and ensuring data trustworthiness is a major driver for success. Although this is closely related to the problem of undeterminable risk, it is often emphasised as integral part of the motivation for adopting trust ensuring solutions. For example, \cite{Alkhelaiwi.2015} argue, that data is useless if it cannot be trusted and that it is a key to success. This is in line with \cite{Ardagna.2021} stating, that system automation is assuming data to be trustworthy, which if failed to meet, results in wrong untrusted services. On top of that, the authors claim the higher the trustworthiness of the data, the higher the decision accuracy. This is further emphasised by \cite{Karthik.2016} which state that trustworthy data is increasing the reliability of applications. \cite{Ardagna.2021} also found, that decisions need to be verifiable in order to provide accountability of e.g. autonomous systems. Therefore they argue, that trustworthy data is necessary to ensure this.

\subsection{Design Objectives}
\begin{figure}[!h]
  \vspace{-0.2cm}
  \centering
      {\includegraphics[width = 7.5cm]{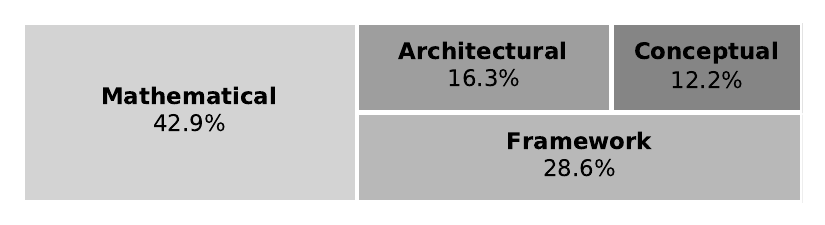}}
  \caption{ The artifact distribution of the 51 investigated artifacts, partially following \cite{Junior.2021}. }
  \label{fig:artifact-distribution}
 \end{figure}

In order to define our design objectives, it is imperative that prior research is investigated, thus the solution space is outlined. Therefore, based on our SLR, we identified 51 previous artifacts concerned with enhancing data trustworthiness.
Building on top of \cite{Junior.2021}, the following four categories were used in order to classify the identified artifacts: conceptual, mathematical, architectural, and framework. The resulting distribution is pictured in Figure~\ref{fig:artifact-distribution}.

We found that although there are many artifacts already, especially in the domain of IoT, most of the existing artifacts were not yet at a stage at which they can be easily adopted, with the majority of papers describing artifacts on a theoretical level. Most artifacts do however have common goals, upon which we based our design objectives. They are described in the following.

\textbf{Decrease Risk.} A central part of the identified problem space are the risks related to using data for which no assumptions about the trustworthiness can be made. This factor has been addressed by a number of existing artifacts. One way of mitigating this is by providing the consumer with enhanced decision making capabilities, for example by explicitly communicating additional information. \cite{Suhail.2018} for instance suggest to provide consumers in an IoT network with the ID of participating nodes, timestamps and the data flow, which were involved in the creation as well the transfer of the data. Based on this, the consumer is enabled to estimate whether the perceived risk of using this data is at an acceptable level. This is in line with \cite{Foidl.2023} which proposed a model in order to compile the quality of data sources, data storage, and the data provider into a trust score ranging from 0 to 1.

Therefore, this way of decreasing risk aims at the consumer to avoid using data which might not be appropriate for their use-case. This approach however requires the consumer to have an in-depth understanding of how different techniques and pieces of information relate to what degree of risk. This is further emphasised by \cite{Haron.2017}, who argue that most trust models leave it up to the user to decide whether the provided trustworthiness assessment, represented either as binary state, or a range of values, is appropriate for them.

Consequently, another way of decreasing risk is by directly employing concrete measurements. This is to ensure, that different risks are mitigated right at the source or before they even reach the consumer. For example, \cite{Tao.2020} propose a secure data collection scheme, which aims to enhance data trustworthiness, by only delivering data which meets a predefined set of quality criteria. Another way of tackling this issue is presented in \cite{HuiLin.2018}. In the study, the authors present a reputation-based mechanism, which aims to defend against internal attacks. Their aim was to increase data trustworthiness by ensuring data veracity.

Based on our findings, we choose to adopt the goal of decreasing risk. We do so, because one major goal of our study is to tackle the risks consumers face when using data, therefore we aim to achieve the same goal as existing artifacts do.

\textbf{Decrease Complexity.} Addressing the problem of complexity, most artifacts share the objective of decreasing the complexity of trust assessment for the consumer. This is the reason why most of the artifacts identified in our SLR opted for computing a trust score, providing an intuitive metric. This is in line with \cite{Jaigirdar.2019}, arguing that e.g. complete provenance graphs are too complex for users to grasp in order to assess the trustworthiness of data. On top of that, the authors found that users need to be able to understand the measurement in order to feel the data is trustworthy.

We, too, adopt this design goal, as it is a central aspect of our target domain of inter-organisational data sharing. Many different users with different backgrounds will need to grow confident in the assessment of data trustworthiness. As a result, they need to be able to understand the given information, so they can base their decisions on it.

\textbf{Enhance Trust.} By far the most important goal existing solutions aim to address is trust, i.e., increasing the confidence of the consumer in the data they use. \cite{Alkhelaiwi.2015} highlight that by arguing that data trustworthiness is essential for users to utilise the data confidently. This is in line with research on inter-organisational data sharing, stating that trust is the most important factor for it to succeed \cite{Tocco.2022}. Consequently, most artifacts address this point by increasing transparency. In \cite{Leteane.2024b} e.g., the authors enhance the trustworthiness of traceability data, by providing tamper-proof evidence of the provenance of data using \textit{blockchain}.

Following this, we adopt this design objective as well, as one major goal of our study is to enhance the overall trust of consumers on a data level. And although the goals of trust and risk seem quite similar, there are some clear distinctions to be made. This is due to the fact that in order to decrease risks, one is mostly concerned with avoiding doing something or employing measurements in order to prevent something to happen. Contrarily, trust in a general sense, is quite subjective to the entity which perceives it and comprised of many different dimensions \cite{Zou.2023}. Therefore, being transparent about the data itself and the measurements in place in order to ensure different criteria is as important as employing them in the first place. Still, both goals are closely related.

\textbf{Enable Interoperability.} One design objective we did not find among the identified artifacts is interoperability. Most of the existing artifacts aim at tackling a very specific case or a broader domain such as IoT. Still, the existing solutions could play a major important role in the context of inter-organisational data sharing, to measure, assess or ensure trustworthiness. This is because many of the trust models and frameworks could be applied to other domains or combined in order to cover the whole data flow in inter-organisational data sharing. Yet, there is no standardised trust model or overarching solution, which aims at utilising the existing technologies to enhance trustworthiness at a data level \cite{Ebrahimi.2022}.

For a similar reason, the identity LoA-based assurance technique eIDAS was introduced by the EU. More specifically, the eIDAS directive mentions the fragmented approach to identity verification and a lack of interoperability among identification techniques as key drivers for passing the regulation \cite{EuropeanParliament.23July2014}.

We therefore believe that the current heterogeneous landscape of existing artifacts can benefit from taking a similar approach. This way, a combination of existing solutions could be utilised to assure data trustworthiness in the challenging environment of inter-organisational data sharing. Hence, we apply and adopt this design goal from the domain of LoAs.

\subsection{Artifact Description}
\begin{figure}[!h]
  \centering
      {\includegraphics[width = 6.5cm]{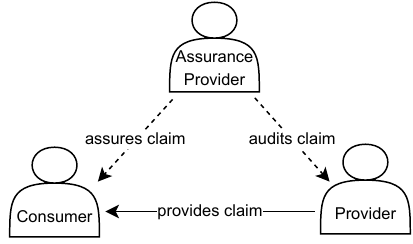}}
  \caption{ Abstract actor model of Data LoA. }
  \label{fig:loa-actors}
 \end{figure}

Based on the derived design knowledge and the defined design goals, the artifact was developed by a cross-functional team of researchers and practitioners in an agile way. The resulting artifact described in the following is an LoA-based assurance technique. We call it \textit{Levels of Assurance for Data Trustworthiness}, or short \textit{Data LoA}. As it is a first-iteration DSR artifact, it specifically concentrates on outlining the overall idea as well as actors and their relationships. This ensures that the most important elements are identified as a first, allowing for incremental additions and modifications. The definition of the specific levels however was not part of our scope in this iteration.

Similar to LoAs within other domains, we define Data LoA as follows: \textit{Levels of Assurances for Data Trustworthiness refer to the degree of confidence which with a dataset's underlying information can be trusted to be actually true.} Thus, it aims at grading the confidence a relying party should put into it, in respect to the risk which potentially arise in case the assurance does not hold true. Moreover, we propose the three following actors as well as their interactions, based on an abstract model pictured in Figure~\ref{fig:loa-actors}.

\subsubsection{Data Consumer}
The data consumer is the relying party, speaking in the sense of LoAs. In the context of the ISO/IEC 15408-1 standard another suitable terminology could be risk owner. It is the party which consumes the data, thus it at risk in case the data trustworthiness claim does not hold true. Therefore, the data consumer utilises the Data LoA in order to decide whether or not to use a certain set of data, based on the related risks.

\subsubsection{Data Provider}
The data provider is - in the sense of LoAs - the claimant. Thus, it is the party which claims that a certain set of data offers a certain degree of trustworthiness. More specifically, by claiming a certain Data LoA the data provider claims that appropriate care was taken in order to reach some degree of confidence about the data's trustworthiness. 

\subsubsection{Assurance Provider} 
The assurance provider is employed in the same sense as in LoAs. It is usually a third-party, which acts as trustworthy auditor and assurer between the data provider and the consumer. Although not strictly necessary, it greatly influences the amount of trust the data consumer can put into the assured claims. Self-asserted claims are usually not trusted to a great extend. Therefore, it can be beneficial to have a trustworthy third-party which audits the measurements taken by the data provider. 

\subsubsection{Interactions}
In order to establish a Data LoA, the following interactions between the three mentioned actors are performed. First off, the data provider needs to create a claim. Then, based on this, the assurance provider needs to conduct an audit of the given claim. To do so, the data provider needs to provide the assurance provider with sufficient evidence, proving that the claim indeed holds true. After that, it is at the discretion of the assurance provider to make an assurance. This can either be in the form of certifications, reports or other ways of communicating proofs.

Using the assured claim in combination with a dataset, provided by the data provider, the data consumer can then choose to put its confidence in the trustworthiness of the data or not. Hence, it is up to the data consumer, whether to believe the assured claim or the assurance provider as a whole.

\subsection{Demonstration \& Evaluation}
Proceeding the DSR approach, we aim at evaluating the artifact through demonstration. Therefore, we follow \cite{Hevner.2004} by opting to conduct an \textit{experimental simulation}, as well as an \textit{informed argument} as part of a descriptive evaluation. According to \cite{Gregor.2013}, the evaluation through a PoC is sufficient for novel artifacts.

Hence, we implement the Data LoA artifact in the domain of data spaces, in order to employ it in a sophisticated environment for inter-organisational data sharing. This allows us to build upon a strong landscape of proven data sharing enabling technologies.

On top of that, our aim is to investigate the impact of our artifact to the domain and the implications it might have for data spaces. Additionally, in this PoC we solely focus on the interaction between provider and consumer in order to reduce complexities. Therefore, we assume a self-asserted claim.

\begin{figure*}[!h]
  \centering
      {\includegraphics[width = 14cm]{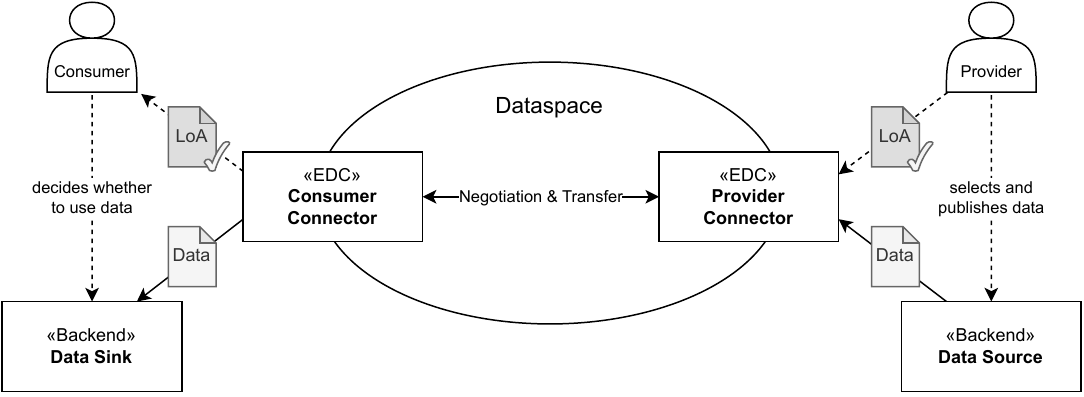}}
  \caption{ Proof of concept for Data LoA, as used in experimental setup. }
  \label{fig:loa-poc}
 \end{figure*}

The PoC reflects a minimal data space. It includes only required parts, in order to showcase and evaluate our artifact. Hence, it comprises the following components, as pictured in Figure~\ref{fig:loa-poc}: \textit{Data Source}, \textit{Data Sink} and \textit{Data Space}.

The Data Space is comprised of two \textit{data space connectors}, each offering all required functionalities for a sovereign data exchange in a data space, both for provider and consumer. Thus, they offer features for data discovery, policy negotiation as well as data transfer. The connectors were implemented using a community-driven open-source framework, namely the \textit{Eclipse Data Space Components}\footnote{\url{https://projects.eclipse.org/projects/technology.edc} [Accessed 30.03.2025]}.

The Data Source and Data Sink naturally resemble a data generating or consuming entity or process, respectively. They both are simple Python backends, aiming to provide or accept data over a REST API.

Furthermore, the PoC is deployed using Docker on a virtual machine running Linux Ubuntu. With everything in place, the following steps are performed: First of, the provider selects a dataset from his data source, which they want to publish to the data space. Upon publishing, an asset gets registered at the provider connector, making the dataset available to the data space. The asset is part of a \textit{data catalog} and describes the dataset as well as the usage policies which are linked to it. Furthermore, the provider can include miscellaneous information in it. Therefore, the provider includes the respective Data LoA claim in the newly created asset as well.

After the asset is successfully made available in the data space, the consumer utilises their connector, in order to request the provider's catalog. Upon retrieval, the consumer is able to inspect the registered asset. Usually, the consumer decides whether or not to request and use the asset, based on the provided information in the catalog. However, the information provided is often times rather sparse, defining only a general description of the data. Yet, in this case, the consumer is provided with an additional Data LoA claim. We assume that the Data LoA claim in this experiment is suitable for the consumer. Therefore, the consumer decides to request and negotiate the data offer. After that, the data is successfully transferred and ready for use for the consumer.

The experimental simulation showcases how the Data LoA claim is presented and communicated between both actors in the context of inter-organisational data sharing. Therefore, it demonstrates how the consumer is first of all provided with details about the data trustworthiness. This approach, is well-suited to increase overall trust by enhancing transparency. Also, only if the degree of confidence of the data trustworthiness claim is feasible for their case, the consumer decides to actually request and use the data. This enables the consumer to greatly decrease risks, by using only data which suits their needs. Furthermore, the overall concept of LoAs is well-suited to decrease complexity for the consumer. Although this simulation lacked a concrete definition of different levels, one can easily imagine that a few levels are easy for the consumer to comprehend.

Furthermore, considering the derived design knowledge, it clearly shows how existing problems are met and most design goals are addressed. Solely the design goal of interoperability could not be demonstrated in this experiment as emphasise was put on enhanced risk management capabilities by providing an easy to comprehend trust-enhancing claim.

\section{Discussion}
\label{sec:discussion}
In this study we designed and proposed a novel artifact for assuring the trustworthiness of data. Our aim was to address the unbalanced view of trust in inter-organisational data sharing. We chose to focus on trust enhancing techniques on a data level. We therefore conducted a DSR approach, building on top of existing data trustworthiness solutions, as well as the commonly adopted technique of LoAs. This resulted in the Data LoA as a first iteration concept. We proceed by discussing the implications of our study, as well as limitations and future work.

First, we identified and emphasised the need for an increased research effort in the domain of consumer trust in inter-organisational data sharing. We highlighted the related risks consumers are prone to, when blindly trusting data. Based on this, this paper investigated existing solutions for measuring and assuring data trustworthiness by conducting a SLR, trying to find suitable methods that fit our domain. We found, that the current solution landscape is quite heterogeneous and lacks a common standardised trust model. Furthermore, we identified the commonly acknowledged challenges of undetermined risks, trustworthiness complexity and demand.

Second, it was demonstrated, how the Data LoA artifact was able to address most of the identified challenges, by integrating it in a data sharing environment. It was showcased, how the Data LoA claim can be communicated and exchanged, enabling the consumer to make informed decisions, addressing the aim of enhanced risk management capabilities, specifically. Furthermore, by providing the consumer with more details related to the trustworthiness of data, transparency and thus the overall consumer trust could be enhanced. On top of that, we adopted a commonly acknowledged goal of LoAs, i.e., decreasing the overall complexity by laying the foundation for LoAs for data trustworthiness. However, concrete levels are yet to be defined and the design objective of interoperability to be addressed.

Lastly, we contribute to the body of design knowledge, using our derived problem and solution space. To the best of our knowledge, this is the first work which explicitly stated design knowledge as well as design objectives. Hence, we achieved to establish a sound foundation to build our artifact upon on. According to \cite{Gregor.2013}, our contributed artifact is classified as an \textit{improvement}, as it is built on top of existing artifacts. We therefore add to the community, by providing an enhancement of existing solutions, as well as design knowledge for future work to build upon on.

\subsection{Limitations \& Future Work}
Despite conducting a rigorous design approach, our study is subject to limitations. First, our design knowledge, acting as foundation for our design activities, was derived using a SLR approach. Naturally, literature reviews are limited by the coverage they achieve. Therefore, in order to increase the coverage and mitigate missing relevant literature, we conducted a back- and forward search as part of the SLR.

Second, our designed artifact is still in an early stage. We therefore didn't specify the different levels of assurance, missing a concrete definition of measurements necessary in order to achieve different degrees of confidence. We chose to do so, as our main purpose of this study was to motivate the overall topic as well as lay the foundation for future work to build upon on. Therefore, we carefully evaluated the artifact in an experimental simulation as PoC, employing it in a conceptual data sharing environment, focusing solely on the interaction between consumer and provider.

Third, while we identified interoperability as one of our design goals, our first iteration artifact mainly focused on achieving the other goals. This was to ensure that our development activities stayed focused on addressing what's commonly agreed as major problems in the field. Still, having adopted the overall technique of LoAs to the domain of data trustworthiness, we are confident that future iterations are capable of addressing this.

Consequently, we identify the following touch points and next steps for future work. First off, more work is needed on the topic of data trustworthiness. As there is currently no commonly accepted notion of data trustworthiness, it is challenging to come up with a proper definition of levels, let alone make them comprehensible for consumers. Therefore, a literature survey may be suitable to address this.

Following this, more work should be put into the Data LoA concept by performing additional DSR cycles. We are confident that the overall idea is capable of addressing the identified challenges sufficiently. Still, as it is a novel concept, there are many points to investigate in detail. First of all, a proper definition of the levels are needed. This is crucial in order for providers to be able to make sound claims, as well as for consumers to conduct a proper risk assessment. To do so, current work on potential risk dimensions and attack taxonomies have to be considered, in order to define the LoAs in a risk-based manner. Also, other dimensions of data trustworthiness such as reputation of source, or country of origin have to be considered. Additionally, the body of design knowledge should be enhanced by, e.g., deriving design principles in order to guide future DSR cycles.

Furthermore, the Data LoA concept itself needs to be looked upon from a greater perspective. Therefore, usage incentives, challenges in implementing and adopting it, as well as domains have to be identified. This is to enable a wide-spread adoption by clearly communicating the addressed domains as well as benefits and trade-offs. For instance, according to \cite{He.2015,Hou.2024}, there are trade-offs to be made when choosing between data trustworthiness and e.g. costs or privacy. Therefore, it is important to weigh the required investments with the expected potential risks and their resulting damages.

For that reason, domains and use-cases have to be identified, in which such a concept could be beneficial. From our perspective, this is the case e.g. in critical infrastructure, such as water or power supply, or other automated systems such as in healthcare. But also AI could greatly benefit from the concept, by e.g. weighing data with low LoA lower, than data with high LoA. This way, less trustworthy data could still be facilitated for training.

\section{Conclusions}
\label{sec:conclusion}
In this study, we have designed a novel LoA for data trustworthiness artifact, aiming at addressing the shortcomings of existing solutions. By doing this, we addressed the unbalanced view of trust from a consumer perspective in inter-organisational data sharing, highlighting the need for trust enhancing solutions. We accomplished this by opting for a DSR approach, building on top of existing research, deriving challenges and goals.

We found, that although our artifact meets most of the identified challenges and goals in theory, more work is needed in order for it to be deployable in a real life scenario. Especially the goal of interoperability needs more attention, as well as the definition of the different levels and the adoption of such. Further DSR cycles should be performed in order to incrementally enhance the artifact.

Our work contributes to the realm of design research by proposing derived design knowledge, as well as a new artifact. We encourage researchers to utilise our findings as foundation in order to further investigate the subject. Finally, we hope our contributions help to increase the research efforts in order to address the shortcomings in regards to consumer trust, ultimately increasing data sharing activities across organisations.

\section*{Acknowledgements}
\subsection{CRediT author statement}
\textbf{Florian Zimmer:} Conceptualisation, Methodology, Software, Validation, Investigation, Writing - Original Draft, Visualisation. \textbf{Janosch Haber:} Conceptualisation, Writing - Review \& Editing, \textbf{Mayuko Kaneko:} Conceptualisation, Writing - Review \& Editing, Project administration.

\bibliographystyle{plainnat}
\bibliography{literature}

\end{document}